\documentclass[]{aa}  
\usepackage{graphicx}
\usepackage{txfonts}
\usepackage{natbib} 

\def\kms{$\mathrm{km\;s}^{-1}$}

\def\msun{M$_{\odot}$}
\def\ms{$M_\bullet-\sigma_c$}

\def\mbh{$M_\bullet$}

\def\nii{[N~{\small II}]}

\def\niig{[N~{\small II}]$\,\lambda6583$}
\def\niipg{[N~{\small II}]$\,\lambda\lambda6548,6583$}
\def\ha{H$\alpha$}
\def\sii{[S~{\small II}]}

\def\siipg{[S~{\small II}]$\,\lambda\lambda6716,6731$}

\def\fdg{\hbox{$.\!\!^\circ$}}
\def\farcm{\hbox{$.\mkern-4mu^\prime$}}
\def\farcs{\hbox{$.\!\!^{\prime\prime}$}}

\begin{document}

\title{The ultraviolet flare at the center of the elliptical galaxy NGC~4278} 


\author{A. Cardullo\inst{1}
        \and E.~M.~Corsini\inst{1}
        \and A.~Beifiori\inst{1}
        \and L.~M.~Buson\inst{2}
        \and E.~Dalla~Bont\`a\inst{1}
        \and L.~Morelli\inst{1}
        \and A.~Pizzella\inst{1}
          \and F.~Bertola\inst{1}
	}

\offprints{A. Cardullo}

\institute{Dipartimento di Astronomia, Universit\`a di Padova, 
  vicolo dell'Osservatorio 2, I-35122 Padova, Italy\\
\and INAF, Osservatorio Astronomico di Padova, 
  vicolo dell'Osservatorio 5, I-35122 Padova, Italy\\
             }

\headnote{Research Note}

\titlerunning{UV flare in NGC 4278}
\authorrunning{Cardullo et al.}

\abstract
{A large fraction of otherwise normal galaxies shows a weak nuclear
  activity. One of the signatures of the low-luminosity active
  galactic nuclei (LLAGNs) is the ultraviolet variability which was
  serendipitously discovered in the center of some low-ionization
  nuclear emission-line region (LINER) galaxies.}
{There is a pressing need to acquire better statistics about UV
  flaring and variability in galaxy nuclei, both in terms of the
  number and monitoring of targets. The Science Data Archive of the
  Hubble Space Telescope was queried to find all the elliptical
  galaxies with UV images obtained in different epochs with the Wide
  Field Planetary Camera 2 (WFPC2) and possibly with nuclear spectra
  obtained with the Space Telescope Imaging Spectrograph (STIS) in the
  region of the \ha\ emission line. These data were found only for the
  elliptical radiogalaxy \object{NGC~4278}.}
{The UV flux of the nuclear source of NGC~4278 was measured by means
  of aperture photometry on the WFPC2/F218W images obtained between
  June 1994 and January 1995. The mass of the central supermassive
  black hole (SBH) was estimated by measuring the broad components of
  the emission lines observed in the STIS/G750M spectrum and assuming
  that the gas is uniformly distributed in a sphere.}
{The nucleus of NGC~4278 hosts a barely resolved but strongly variable
  UV source. Its UV luminosity increased by a factor of 1.6 in a
  period of 6 months. The amplitude and scale time of the UV flare in
  NGC~4278 are remarkably similar to those of the brightest UV nuclear
  transients which were earlier found in other LLAGNs.  The mass of
  the SBH was found to be in the range between $7\,\times\,10^7$ and
  $\,2\,\times\,10^9$ \msun .  This is in agreement with previous
  findings based on different assumptions about the gas distribution
  and with the predictions based on the galaxy velocity dispersion.}
{All the LINER nuclei with available multi-epoch UV observations and a
  detected radio core are characterized by a UV variable source. This
  supports the idea that the UV variability is a relatively common
  phenomenon in galaxy centers, perhaps providing the missing link
  between LINERs and true AGN activity.}

\keywords{Galaxies: active --- Galaxies: elliptical and lenticular, cD
  --- Galaxies: individual: NGC 4278 --- Galaxies: nuclei ---
  Ultraviolet: galaxies --- Black hole physics}

\maketitle

\section{Introduction}
\label{sec:introduction}

A large fraction of otherwise normal galaxies shows a weak nuclear
activity. These low-luminosity active galactic nuclei (LLAGNs) occupy
the faintest end of the luminosity function of the AGNs and have very
low accretion rates or radiative efficiencies onto the central
supermassive black hole \citep[SBH, see][for a review]{Ho2008}.

One of the signatures of the LLAGNs is the ultraviolet (UV)
variability observed with Hubble Space Telescope (HST) in the nuclei
of some low-ionization nuclear emission-line region (LINER) galaxies.
There have been a number of reports after the serendipitous discovery
of an UV flare in the center of the elliptical galaxy
\object{NGC~4552} \citep{Renzini1995, Cappellari1999}. The images of
its nucleus obtained in a five years period with the Faint Object
Camera (FOC) showed an increase of the UV luminosity by a factor of
4.5 followed by a dimming of a factor 2.
\citet{OConnell2005} unveiled quite a similar phenomenon in the giant
elliptical \object{NGC~1399} using the Space Telescope Imaging Spectrograph
(STIS), while a rapidly fading UV source was detected in the Virgo
cluster spiral \object{NGC~4579} \citep{Maoz1995, Barth1996}.
This kind of research received a more systematic approach by
\citet{Maoz2005}, who monitored the UV variability of a sample of 17
LINER galaxies with compact nuclear UV sources by means of the HST
Advanced Camera for Surveys (ACS). They detected a significant UV
variability in almost all the sample galaxies, which were mostly
spirals, with amplitudes ranging from few to $50$ percent.

This suggests that the UV variability is a relatively common
phenomenon in galaxy centers. Given to this, there is a pressing need
to acquire better statistics, both in terms of the number of targets
and in terms of monitoring the UV-variable nuclei.
\citet{Maoz1995} already queried the HST Science Data Archive for ACS
data and observed mostly spiral galaxies. We searched the Wide Field
Planetary Camera 2 (WFPC2) archive for all the RC3
\citep{deVaucouleurs1991} elliptical galaxies
having UV images obtained with the same filter in different epochs
before 2008. We found 37 objects \citep[12 LINERs, 2 Seyferts, 1
  transition and 1 H~{\small II} nucleus, 17 quiescent galaxies, and 4
  objects with unknown nuclear activity;][and NASA/IPAC Extragalactic
  Database]{Ho1997} with at least one F218W or F300W image. NGC~4278 was the
only galaxy with multi-epoch WFPC2/F218W images, which was not studied
before.  A nuclear spectrum obtained within a subarcsecond aperture by
STIS in the \ha\ region was also available in HST Science Data
Archive.
In this paper, we present and discuss the results about the UV
variability of its nucleus including the estimate of the mass of its
SBH based on STIS spectroscopy.

%
\begin{figure} 
\centering
\includegraphics[width=9cm]{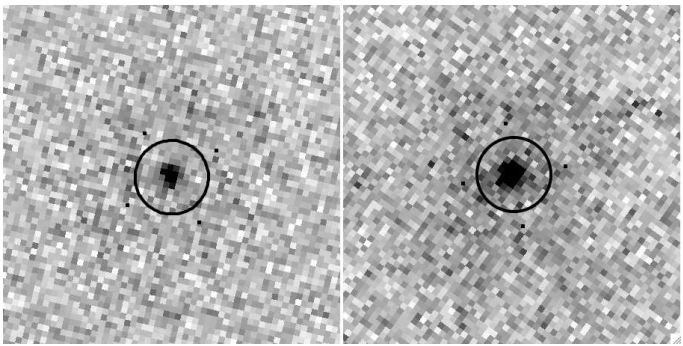}
\caption{WFPC2/F218W images of the NGC~4278 nucleus of 2 June 1994
  (left) and 14 January 1995 (right), plotted with the same grayscale.
  Each panel is $2\farcs5$ on a side, north is up and east at the
  left.  The UV flux was measured within the aperture marked by the
  circle ($r=0\farcs28$).}
\label{fig:aperture}
\end{figure}

\section{NGC~4278}
\label{sec:n4278}

NGC~4278 is a large ($4\farcm1 \times 3\farcm8$, RC3) and bright
($B_T=11.09$, RC3) elliptical galaxy. It is classified E1--2 and hosts
a LINER nucleus \citep[L1.9, ][]{Ho1997}.  Its total absolute
magnitude is $M^0_{B_T}=-20.03$ (RC3), adopting a distance of 16.1 Mpc
\citep{Tonry2001}. NGC~4278 is member of the Coma I cloud
\citep{Forbes1996}.
It was one of the first elliptical galaxies in which neutral hydrogen
was detected \citep{Gallagher1977} and used to infer the dark mater content
at large radii \citep{Bertola1993}. It is distributed in an inclined
outer ring \citep{Raimond1981}, which is possibly associated to the
inner disk of ionized gas \citep{Goudfrooij1994,Sarzi2006}. The northwest side
of the galaxy is heavily obscured by large-scale dust patches and
filaments which seem to spiral down into the nucleus
\citep{Carollo1997}.

The optical and radio properties of the nucleus have been investigated
in detail \citep[see][and references therein]{Giroletti2005}. The
radio data reveal two symmetric steep-spectrum jets on sub parsec
scale. They emerge from a flat-spectrum core and are responsible for
the bulk of the emission at radio to optical frequencies in a similar
way to that seen in more powerful radio loud AGNs. However, the total
radio luminosity of NGC~4278 is at least 2 orders of magnitude less
than those objects \citep{Condon1998}. This makes NGC~4278 a LLAGN
\citep{Giroletti2005}.

\section{Observations, data reduction, and analysis}
\label{sec:uv}

\subsection{Nuclear ultraviolet variability}
\label{sec:wfpc2}

The multi-epoch images obtained with the WFPC2 and the F218W filter
were retrieved from the HST Science Data Archive. A 1800 s exposure
was taken on 2 June 1994 (Prog. Id. 5380, P.I. A. Koratkar). Two
exposures of 2200 s and 2300 s were obtained on 14 January 1995
(Prog. Id. 5381, P.I. A. Koratkar).
The exposures were taken with the telescope guiding in fine lock,
which typically gave an rms tracking error of $0\farcs003$. We focused
our attention on the Planetary Camera (PC) chip where the nucleus of
the galaxy was centered. This consists of $800 \times 800$ pixels of
$0\farcs0455 \times 0\farcs0455$ each, yielding a field of view of
about $36'' \times 36''$.

The images were calibrated using the standard WFPC2 reduction pipeline
maintained by the Space Telescope Science Institute (STScI).
Reduction steps including bias subtraction, dark current subtraction,
and flat-fielding are described in detail in the WFPC2 instrument and
data handbooks \citep{Holtzman1995b, Baggett2002, McMaster2008}.
Subsequent reduction was completed using standard tasks in the STSDAS
package of IRAF\footnote{Imaging Reduction and Analysis Facilities is
  distributed by National Optical Astronomy Observatories
  (NOAO).}. The bad pixels were corrected by means of a linear
one-dimensional interpolation using the data quality files and the
WFIXUP task. The two 1995 images were aligned and combined using
IMSHIFT and knowledge of the offset shifts. Cosmic ray events and
residual hot pixels were removed using the LACOS\_IMA procedure
\citep{vanDokkum2001}.  The cosmic ray removal and bad pixel
correction were checked by inspection of the residual image between
the cleaned and the original frame to ensure that the nuclear region
was not affected by light loss. The residual cosmic rays and bad pixels in
the PC were corrected by manually editing the resulting image with
IMEDIT. The sky level ($\sim1$ count pixel$^{-1}$) was determined from
apparently empty regions in the Wide Field chips and subtracted from
the PC frame after appropriate scaling.

The flux calibration was performed by adopting the Vega magnitude
system \citep{Whitmore1995} and by taking into account for the time
dependence of the UV response \citep{McMaster2002}.
In fact, the presence of contaminants within WFPC2 causes a gradual
build-up of material on the cold CCD faceplate of the camera,
resulting in a decrease in the UV throughput. The contaminants are
evaporated by periodically heating the camera in order to restore the
instrumental throughput to its nominal value. The contamination rate
is remarkably constant during each decontamination cycle, and can be
accurately modeled by a simple linear decline following the
decontaminations. The observed fluxes were corrected by assuming a
decline in the F218W/PC normalized count rate of
$(4.78\pm0.28)\times10^{-3}$ per day since decontamination. This was
derived during the decontamination cycles performed between April 1994
and June 1995 \citep{McMaster2002}, which nicely bracket the
observations of NGC~4278.

\begin{figure} 
\centering
\includegraphics[width=9cm]{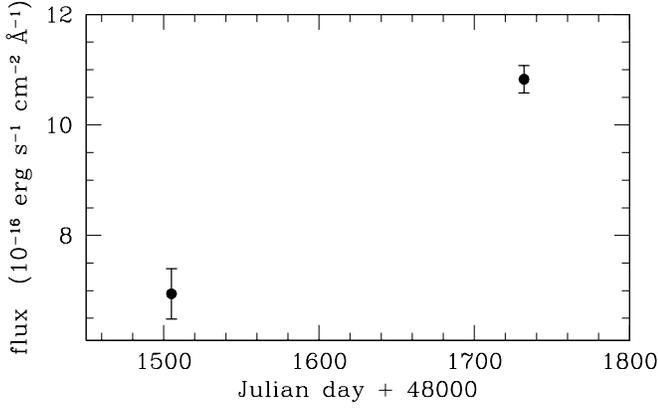}
\caption{UV light curve in WFPC2/F218W band for the nucleus of
  NGC~4278.  Points correspond to fluxes measured on 2 June 1994 and
  14 January 1995, respectively.}
\label{fig:flare}
\end{figure}

Evidence of the presence of a nuclear source was found in the two
final images of NGC~4278. It is barely resolved ($\rm
FWHM\,=\,0\farcs07$) when compared to the WFPC2/F218W point spread
function (PSF, $\rm FWHM\,=\,0\farcs06$) derived with the TINY TIM
package \citep{Krist1999}.
The total flux of the nuclear source was estimated as the flux in the
circular aperture of radius of $0\farcs28$
(Fig. \ref{fig:aperture}). The background level was determined as the
median of the flux within the annulus of $0\farcs37$--$0\farcs51$
centered on the source. The correction for the finite aperture radius
has been done by multiplying counts by $1.18\pm0.01$, based on the
encircled-energy value of a point source tabulated in
\citet{Holtzman1995a}. The errors were calculated taking in account
the Poisson and CCD readout noises, charge transfer efficiency,
correction for contamination, and correction for finite aperture.
The UV flux of the central source of NGC~4278 in the WFPC2/F218W
passband increased from $(6.94\pm0.46)\times 10^{-16}$ to
$(10.83\pm0.25)\times 10^{-16}$ erg cm$^{-2}$ s$^{-1}$ \AA$^{-1}$ from
2 June 1994 to 14 January 1995 (Fig. \ref{fig:flare}).

\subsection{Estimate of the mass of the central black hole}
\label{sec:stis}

The long-slit spectrum of the nucleus of NGC~4278 obtained with
STIS in the region of the \niipg, \ha, and \siipg\ emission lines
(Prog. Id. 7403, P.I. A. Filippenko) was retrieved from the HST Science
Data Archive.  The G750M grating was used at the secondary tilt
$\lambda_{\rm c}\,=\,6581$ \AA\ covering the wavelength range
6295--6867 \AA. The spectrum was taken with the
$0\farcs2\,\times\,52''$ slit placed across the galaxy nucleus with a
position angle of $87\fdg9$. The total exposure time was 3128 s. The
dispersion is $0.55$ \AA\ pixel$^{-1}$.  The instrumental resolution
is 1.6 \AA\ (FWHM) corresponding to $\sigma_{\rm inst}\approx 32$
\kms\ at \ha. The spatial scale of the $1024\times1024$ SITe CCD is
$0\farcs05$ pixel$^{-1}$.

The spectrum was reduced using IRAF and the STIS reduction pipeline
maintained by the STScI. The basic reduction steps including overscan
subtraction, bias subtraction, dark subtraction, and flatfield
correction are described in detail in the STIS instrument and data
handbooks \citep{Kim2007, Dressel2007}. The cosmic ray events and hot
pixels were removed using the task LACOS\_SPEC by
\citet{vanDokkum2001}. The residual bad pixels were corrected by means of
a linear one-dimensional interpolation using the data quality
files. The wavelength and flux calibration as well as geometrical
correction for two-dimensional distortion were performed following the
standard STIS reduction pipeline and applying the X2D task. This task
corrected the wavelength scale to the heliocentric frame as well.

\begin{figure}
\centering
\includegraphics[width=9cm]{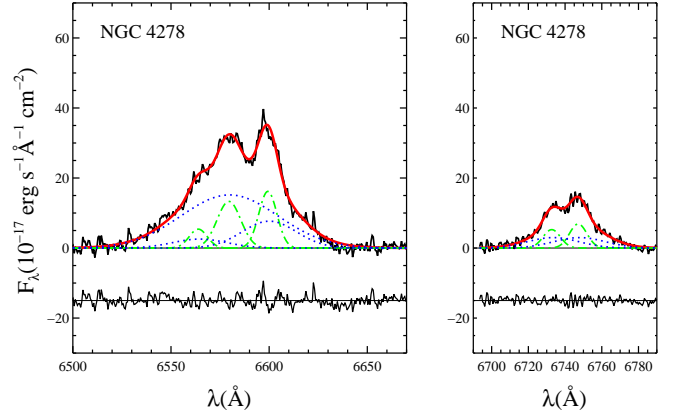}
\caption{Continuum-subtracted central spectrum of NGC~4278 in the
  \ha+\nii\ (left panel) and \sii\ (right panel) regions. In each
  panel, the red solid line shows the overall line blend, whereas the
  green dashed-dotted lines and blue dotted lines show the
  adopted narrow and broad Gaussian components, respectively. 
  Shown are also the fit residuals, offset for better visibility.}
\label{fig:spectrum}
\end{figure}

The central wavelengths, FWHMs, and intensities of the all the
observed emission lines were measured following \citet{Beifiori2009}.
The broad and narrow components of the emission lines were fitted with
multiple Gaussians, while describing the stellar continuum with a
low-order polynomial. A flux ratio of 1:2.96 was assumed for the
\nii\ doublet, as dictated by atomic physics
\citep[e.g.,][]{Osterbrock1989}, both the \nii\ and \sii\ doublets were
assumed to share a common line centroid and width.
A broad component was needed to describe the
\ha\ emission out to $0\farcs20$ from the
center. The forbidden \nii\ and \sii\ lines required
a broad component from $-0\farcs05$ to $+0\farcs1$.
  Fig.~\ref{fig:spectrum} shows the continuum-subtracted
nuclear spectrum of NGC~4278 with the fitted emission lines. It was
extracted from the three central rows of \ha\ STIS/G750M spectrum
centered on the continuum peak. It
thus consists of the central emission convolved with the STIS spatial
PSF and sampled over a nearly square aperture of $0\farcs15 \times
0\farcs2$ (corresponding to $12\times16$ pc$^2$).

The mass \mbh\ of the central SBH can be estimated from the virial
theorem by assuming that the gas is uniformly distributed within a
sphere of radius $R$ and moves around the SBH with a mean velocity
measured from the width of the broad component of the \ha\ line.

The lower limit on the size of the broad-line emitting region is given
by the upper limit to the gas density as set by the forbidden lines
and \ha\ luminosity. Following \citet{Osterbrock1989}

\begin{equation}
R_{\rm min} = \left( \frac{3 L_{\rm H\alpha}}{4 \pi f n_{\rm e}^2
  \alpha_{\rm H\alpha}^{\rm eff} h \nu_{\rm H\alpha}}
\right)^\frac{1}{3}
\label{equ:vol}
\end{equation}

\noindent where $L_{\rm H\alpha}$ is the total luminosity of the
\ha\ broad component, $f$ is the volume filling factor, $n_{\rm e}$ is
the electron density of the gas, $\alpha_{\rm H\alpha}^{\rm eff}$ is
the recombination coefficient for \ha, $h$ is the Planck constant, and
$\nu_{\rm H\alpha}$ is the frequency of the \ha\ line.
The ratio of the \sii\ lines sets $n_{\rm e} \approx 10^5$ cm$^{-3}$,
whereas the spherical and uniform distribution of the gas gives a
filling factor equal to 1. It is $L_{\rm H\alpha} = 4.13 \times
10^{38}$ erg s$^{-1}$ from the measured flux $(1.34 \pm 0.10)\times
10^{-14}$ erg s$^{-1}$ cm$^{-2}$ and assumed distance. It results
$R_{\rm min}=0.1$ pc. In this region $\sigma_V = 1224\,\pm\,40$ \kms,
which corresponds to a lower limit on the SBH mass of \mbh$\geq
R_{\rm min} \sigma_V^2 /G \simeq 7\,\times\,10^7$ \msun .

The upper limit on the size of the broad-line emitting region $R_{\rm
  max}$ can be estimated from the intrinsic emissivity distribution of
the gaseous sphere. An intrinsically Gaussian flux profile centered on
the stellar nucleus was assumed. It has a $\rm FWHM\,=\,0\farcs19$ (15
pc) when accounting for the STIS PSF ($\rm FWHM\,=\,0\farcs09$). The
choice of a Gaussian parametrization is also conservative, since
cuspier functions would have led us to estimate smaller \mbh . It
results $R_{\rm max}=5.9$ pc and $\sigma_V = 1127\,\pm\,30$
\kms\ which translates into an upper limit on the SBH mass of
\mbh$\leq R_{\rm max} \sigma_V^2 /G \simeq 2\,\times\,10^9$ \msun .

\section{Conclusions}
\label{sec:conclusions}
UV variability with amplitudes ranging from few to 50 percent over a
timescale of a decade was detected in most of the LINER nuclei
observed more than once \citep{Maoz2005} suggesting a possible link
between UV flares and SBH-related activity in LLAGNs.
To acquire better statistics, both in terms of the number of targets and their
monitoring we queried the HST Science Archive for all the elliptical
galaxies with available UV images obtained with the WFPC2 in different
epochs.

Multi-epoch images were found only for NGC~4278, a nearby
radiogalaxy known to host a LLAGN \citep{Giroletti2005}.
It is characterized by a barely resolved nuclear source, which
increased its UV luminosity by a factor of 1.6 in a period of 6
months. The amplitude and scale time of the variation are similar to
those of the UV-brightest nuclear transients, which were earlier
discovered in NGC~4552 \citep{Renzini1995, Cappellari1999}, NGC~4579
\citep{Maoz1995, Barth1996}, and NGC~1399 \citep{OConnell2005}.

These serendipituous findings support the idea the UV variability is a
common event at the center of galaxies where SBHs reside. 
Some alternatives to the AGN interpretation were explored to explain the UV
variability.
\citet{Maoz2005} pointed out that individual supergiants
in galactic nuclei are not plausible candidates
for producing the observed UV flux variations. On the other hand, the
even more brighter Wolf-Rayet and luminous blue variable stars could
only explain the variations measured in the nuclei of lower UV
luminosity.
The fall back of debris onto the SBH, and collisions between
precessing debris orbits \citep{Rees1988, Kim1999} are expected to
produce bright UV/X-rays flares \citep{Ulmer1999}. But given their
rarity \citep[$\approx10^{-4}$ yr$^{-1}$ per galaxy,][]{Magorrian1999,
  Wang2004}, the stellar tidal disruptions can emerge only in all-sky
deep X-ray \citep{Donley2002, Esquej2007} and UV surveys
\citep{Gezari2006,Gezari2008, Gezari2009} and can not account all the
observed variable nuclei. This is particularly true for the galaxies
with repeated episodes of UV variability, like NGC~4552
\citep{Renzini1995, Cappellari1999, Maoz2005}.

It was possible to estimate the mass of the SBH at the center of
NGC~4278. The central width and radial extension of the broad
components of the emission lines were measured over a subarcsecond
aperture in the available STIS spectrum of the nucleus. If the gas is
uniformly distributed within a sphere then it is $7\,\times\,10^7\,\leq$
\mbh\ $\leq\,2\,\times\,10^9$ \msun . This is consistent with the
predictions of the \ms\ relations by \citet{Ferrarese2005} and
\citet{Lauer2007} when adopting a central velocity dispersion
$\sigma_{\rm c}=333\pm8$ \kms\ \citep{Beifiori2009}. There is also a
good agreement with the upper limits on \mbh\ given by
\citet{Beifiori2009}. They measured the nuclear width of the narrow
component of the \niig\ emission line and modelled it as due to the
Keplerian motion of a thin disk of ionized gas around the putative
SBH.  They found \mbh$\,\leq\,5\,\times\,10^7$ for a nearly edge-on
disk and \mbh$\,\leq\,2\,\times\,10^8$ \msun\ for a nearly face-on
disk.

According to \citet{Giroletti2005} this SBH is active and able to
produce the relativistic jets, which are responsible for most of the
emission at optical and radio frequencies of this LLAGN. It is the
same process of the ordinary radio loud AGNs despite a much lower
power.  The AGN interpretation is a promising way to explain the UV
variability. In fact, all the LINER nuclei with multi-epoch
observations and a detected radio core are characterized by UV
variable source \citep{Maoz2005}. This is the case of NGC~4278 too,
suggesting that UV variability could provide the missing link
between LINERS and true AGN activity.
Unfortunately, it is the only elliptical galaxy observed by HST in
different epochs in the UV and which was not studied before. This fact does
not allow us to derive any firm statistical conclusion about the frequency of UV flares at
the center of elliptical galaxies. Nevertheless, it has to be remarked
that out of 37 galaxies the only object with multi-epoch UV observations
turned out to be variable. This is a further suggestion that this
phenomenon has to be quite common.
Additional imaging with the Wide Field Camera 3 recently installed on
HST to monitor the UV variablity in a statistically significant sample
of quiescent and active nuclei and STIS spectra to measure their SBHs
are highly desirable to gain insights on this subject in the next future.

\begin{acknowledgements} 
AC acknowledges the Space Telescope Science Institute for hospility
while this work was in progress. We thank Massimo Stiavelli and
Michele Cappellari for useful discussions and their helpful
comments. This work was made possible through grants CPDA068415 and
CPDA089220 by Padua University.  This research has made use of the
Lyon-Meudon Extragalactic Database (LEDA) and NASA/IPAC Extragalactic
Database (NED).

\end{acknowledgements}


\bibliographystyle{aa}

\end{document}